\documentclass[prb,twocolumn,superscriptaddress]{revtex4-2}
\usepackage[utf8]{inputenc}
\usepackage[american,]{babel}
\usepackage[T1]{fontenc}
\usepackage[pdftex]{graphicx}  
\usepackage{graphicx,xcolor}
\usepackage{dcolumn}
\usepackage{bm}
\usepackage{dsfont}
\usepackage{bbold}
\usepackage{amsmath,amsthm,amssymb}
\usepackage{hyperref}
\usepackage{balance}
\hypersetup{colorlinks,bookmarksopen,bookmarksnumbered,
    citecolor=red,
    linkcolor=red,
    pdfstartview=false,
    urlcolor=red}
\usepackage{braket}
\usepackage{wrapfig}
\usepackage{mathtools}
\usepackage{float}
\usepackage{tabularx}
\usepackage{physics}

\begin{document}
\title{Measurement-induced crossover in quantum first-detection times}

\author{Giovanni Di Fresco}
\affiliation{SISSA-International School for Advanced Studies, via Bonomea 265, 34136 Trieste, Italy}
\affiliation{INFN, Sezione di Trieste, via Bonomea 265, 34136 Trieste, Italy}

\author{Aldo Coraggio}
\affiliation{SISSA-International School for Advanced Studies, via Bonomea 265, 34136 Trieste, Italy}
\affiliation{INFN, Sezione di Trieste, via Bonomea 265, 34136 Trieste, Italy}

\author{Alessandro Silva}
\affiliation{SISSA-International School for Advanced Studies, via Bonomea 265, 34136 Trieste, Italy}

\author{Andrea Gambassi}
\affiliation{SISSA-International School for Advanced Studies, via Bonomea 265, 34136 Trieste, Italy}
\affiliation{INFN, Sezione di Trieste, via Bonomea 265, 34136 Trieste, Italy}

\date{\today}

\begin{abstract}
The quantum first-detection problem concerns the statistics of the time at which a system, subject to repeated measurements, is  observed in a prescribed target state for the first time. 
Unlike its classical counterpart, the measurement back action intrinsic to quantum mechanics
may profoundly alter the system dynamics.
Here we show that it induces a distinct change in the statistics of the first-detection time. For a quantum particle in one spatial dimension subject to stroboscopic measurements, we observe an algebraic decay of the probability of the first-detection time if the particle is free, an exponential decay in the presence of a confining potential, and a time-dependent crossover between these behaviors if the particle is partially confined. This crossover reflects the purely quantum nature of the detection process, which fundamentally distinguishes it from the  first-passage problem in classical systems. 
\end{abstract}
\maketitle

\emph{Introduction.---} 
When does a random process cross an assigned boundary for the first time? This seemingly simple question --- the problem of the first passage time --- is relevant for such diverse phenomena as chemical reactions, neuronal firing, and market fluctuations~\cite{Redner2001, Hughes95,Bray2013,Benichou2014, Metzler2014}, with its earliest formulation tracing back to Schr\"odinger’s work on Brownian motion with absorbing boundaries~\cite{Schrodinger1915}. 
In classical systems, the first-passage statistics provides information on the properties of the noise, on the underlying geometry and on possible interactions. A classical agent is, however, assumed to be under observation at all times. It is not so for quantum systems, where quantum noise, interference and measurement back action are important additional ingredients to be taken into account~\cite{Ambainis2001}.
Indeed, in the quantum realm, the very notion of ``passage'' becomes inseparable from the measurement protocol that defines it, as the observation is an integral part of the dynamics. 
A natural generalization of the classical notion of first passage time is thus the time at which a quantum system, initialized in some state, is first detected in a specified target state. The statistics of this first-detection time in quantum systems with stroboscopic measurements has recently attracted considerable interest both with constant~\cite{Dhar2015-1,Dhar2015QuantumTimeOfArrival, Friedman_2017,Walter_2025, Thiel2105,Yin2019, Yin2025,heine2025quantumwalkshittingtimes,Dittel2023, ryan2025, Walter_2025, Dubey2021, Lahiri2019} or variable~\cite{delvecchio2025optimaldetectionquantumstates} intervals between subsequent measurements.
On the one hand, it was clarified that, for a free particle hopping on a lattice, the probability of being detected at a specific lattice site after $n$ measurements decays as $n^{-3}$ at late times, in contrast to the classical decay $\propto n^{-3/2}$.
On the other hand, it is clear that the quantum first-detection problem differs fundamentally from its classical counterpart 
because the interaction with a measuring device unavoidably perturbs the system evolution. 
This raises a simple yet fundamental question: can we identify features of the first-detection statistics that arise specifically from the system drastic alteration due to measurement back action?

\begin{figure}[h!]
    \centering
    \includegraphics[width=0.65\linewidth]{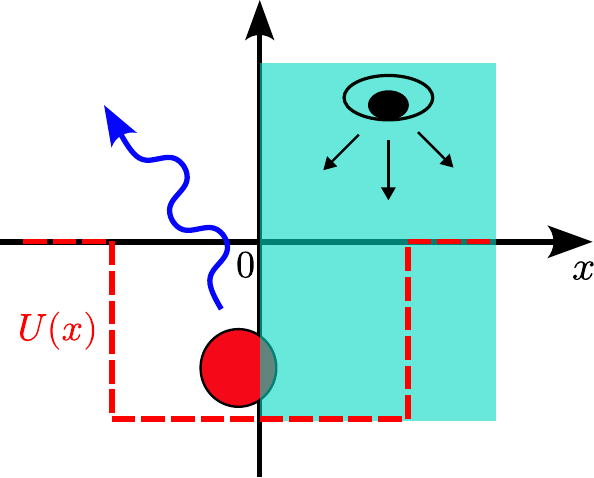}
    \caption{A particle (red circle) in a 
    potential well $U(x)$ is subject to stroboscopic 
measurements that test whether it is in the positive half-line $x>0$ (colored background). The 
measurement back action injects energy into the system, thereby enhancing 
the probability that the particle escapes from the well (wiggly arrow). This, in turn, leads 
to a crossover in the statistics of the first-detection time.}
    \label{fig:1}
\end{figure}

To address this question, we consider a simple detection problem: A quantum particle with coordinate $x$ in one spatial dimension is initially localized at $x=x_0\leq 0$, is subject to a potential $U(x)$ and is monitored at regular time intervals to determine whether it is within the detection region $x > 0$. In particular, we are interested in the statistics of the time at which the first successful detection of the particle occurs.
We first show that, depending on the discrete or continuous nature of the energy spectrum, the asymptotic decay of the probability of the first-detection time at long times is either exponential or algebraic.
We show this by considering, first, the case $U=0$ of a free particle, for which the first-detection probability turns out to decay algebraically, 
and, second, that of a  particle fully confined in a parabolic potential, for which the decay turns out to be exponential. 
We then show that, in intermediate situations in which the particle is confined by a potential well $U$ of finite depth, a distinct and rather uncommon crossover from an exponential to an algebraic decay of the first-detection probability is observed. 
This phenomenon occurs because all measurements, irrespective of their outcome,
inject energy in the system (and influence its energy statistics~\cite{malakar2025}),
allowing the particle to escape confinement and populate the continuum part of its spectrum, thereby reshaping the long-time statistics of the first-detection time.

\emph{Model and measurement protocol.---} The dynamical protocol of stroboscopically monitored quantum particle is the following
\cite{Friedman_2017}: start from an initial state $\ket{\psi(0)}$ and let it evolve for a time $\tau$ according to $\ket{\psi(\tau)} = e^{-iH\tau}\ket{\psi(0)}$, where $H$ is the Hamiltonian of the
system. 
After this evolution, we ask whether the system is in a specific subset of its Hilbert space. For example, one can consider a generic state belonging to this subspace as the target state $\ket{\psi_{\text{t}}}$. To test this, one performs a projective measurement onto 
$\ket{\psi_{\text{t}}}$,
which is successful with probability $p_0 =|\bra{\psi_{\rm{t}}}\psi(\tau)\rangle|^2$. After the measurement, there are two possibilities: either the system is in $\ket{\psi_{\text{t}}}$ 
and the dynamics stops, or it continues. Focusing on the latter case, the state $\ket{\psi(\tau^+)}$ immediately after the measurement is
\begin{equation}
    \ket{\psi(\tau^+)} = \frac{\left(1-D\right)\ket{\psi(\tau)}}
    {\sqrt{1 - p_0}}\label{eq:proj_state}
\end{equation}
where $D = \ketbra{\psi_{\text{t}}}{\psi_{\text{t}}}$. Repeating the measurement $n$ subsequent times, separated by a time $\tau$, one can ask: what is the probability $F_n$ of detecting the particle at the $n$-th attempt after $n-1$ unsuccessful detections? In terms of the (unnormalized) state 
right before the $n$-th measurement~\cite{Friedman_2017} 
\begin{equation}
       \ket{\psi(n\tau^-)} = {e^{-iH\tau}\left[\left(1-D\right)e^{-iH\tau} \right]^{n-1}}\ket{\psi(0)}, \label{eq:n-evolution}
\end{equation}
it is easy to see that~\cite{Friedman_2017}
\begin{equation}
    F_n = \bra{\psi(n\tau)}D\ket{\psi(n\tau)}.\label{eq:Fn_3}
\end{equation}
Let us now use these expressions to investigate the statistics of the first detection time in the positive half-line for a quantum particle in one spatial dimension, as illustrated in Fig.~\ref{fig:1}.
In particular, starting from a single-particle wavefunction $\psi(x,t)$ localized around the position $x= x_0$, the system is probed at stroboscopic intervals of duration $\tau$ to determine whether the particle is within 
the region $x>0$. 
Equation~\eqref{eq:Fn_3} then becomes 
\begin{equation}
    F_n = \int_{0}^{\infty} \!\!\dd x \,|\psi(x, n\tau)|^2 ,
    \label{eq:norm}
\end{equation}
where $\psi(x, n\tau) = \braket{x}{\psi(n\tau)}$. 
If the particle is not detected in the positive half-line, the wavefunction is truncated by setting its amplitude to zero for $x>0$, which corresponds to applying the projector $1-D = \theta(-x)$~\cite{supplementary}, where $\theta(x)$ is the Heaviside step function. 

\emph{Free particle.---} The statistics of the first detection time for a single free particle has been studied previously in a variety of settings, for example in Refs.~\cite{Dhar2015-1, Friedman_2017, Dubey2021}. Here, we consider a particle on the continuum, with the initial condition given by a Gaussian wavefunction centered at $x_0 < 0$ with variance $\sigma$, i.e.,
\begin{equation}
    \psi(x, 0) = \braket{x}{\psi(0)} = \frac{e^{-(x-x_0)^2/(4\sigma^2)}}{(2\pi\sigma^2)^{1/4}}.\label{eq:iintial_free}
\end{equation}
In the time interval between subsequent measurements, the particle evolves with the free Hamiltonian 
\begin{equation}
    H = p^2 / 2.    
\end{equation}
In this context, two regimes naturally arise: (a) the Zeno regime which dominates for $\tau \leq \tau_Z$, where $\tau_Z$ is the Zeno time defined as~\cite{Scardicchio_2001, Facchi_2008}
\begin{equation}
 \tau_Z^{-2} = \braket{\psi}{H^2|\psi} -\braket{\psi}{H|\psi}^2
 \label{eq:def-tauZ}
\end{equation}
and (b) the non-Zeno regime with  
$\tau > \tau_Z$.
It is well established that, in the Zeno regime, the following approximation holds~\cite{Allcock1969-1,Allcokc1969-2, Allcock1969-3, Echanobe2008, Halliwel2009, Halliwell2010}:
\begin{equation}
\begin{split}
    \underbrace{e^{-iH\tau}(1-D)}_n\underbrace{e^{-iH\tau}(1-D)}_{n-1}\underbrace{\ldots}_{\ldots}  \underbrace{e^{-iH\tau}(1-D)}_1 \\
    \qquad \simeq e^{\left[-iH - V_0\theta(x)\right]n\tau},
\end{split}
\label{eq:mapping}
\end{equation}
which maps the effects of a sequence of $n$ measurements onto an effective evolution under a spatially constant complex potential $-iV_0$ within the detection region $x>0$. 
In particular\textbf{},
\begin{equation}
    V_0 \simeq 1/\tau.
\end{equation}
This mapping allows an analytical solution of the problem in the Zeno regime. 
Choosing $\psi(x,t=0)$ as in Eq.~\eqref{eq:iintial_free} with $x_0<0$, the dynamics can be formulated as a scattering problem from a constant non-Hermitian potential. 
The corresponding reflection and transmission coefficients turn out to be \cite{supplementary},  respectively, 
\begin{equation}
    R(k)= \frac{k-q(k)}{k+q(k)} \quad\mbox{and}\quad T(k) = \frac{2k}{k+q(k)},\label{eq:ref_tran}
\end{equation}
where $q(k) = \sqrt{k^2 + 2iV_0}$, and the time-dependent wavefunction reads
\begin{equation}
    \psi(x,n\tau)\!\!= \!\!\left\{\begin{split}&\!\!\!
                \int \!\!\frac{\dd k}{\sqrt{2\pi}}\widetilde{\psi}(k)\left[ e^{ikx} + R(k)e^{-ikx}\right] e^{-ik^2n\tau/2}, \; x<0,\\
                 &\!\!\int \!\!\frac{\dd k}{\sqrt{2\pi}}\widetilde{\psi}(k)T(k)e^{-iq(k)x}e^{-ik^2n\tau/2}, \; x>0,
    \end{split}\right.\label{eq:wave_function_time}
\end{equation}
where $\widetilde{\psi}(k)$ is the Fourier transform of the initial state.
Using Eq.~\eqref{eq:wave_function_time} for $x>0$ in Eq.~\eqref{eq:norm}, it is easy to show that, for large $n$, $F_n$ decays as~\cite{supplementary}
\begin{equation}
    F_n = A(x_0, \sigma) n^{-3},
    \label{eq:Fn_zeno}
\end{equation}
where $A(x_0, \sigma)$ is determined in Ref.~\cite{supplementary}. 
In the Zeno regime, for a detection at point, a similar scaling was also found in the continuum in Ref.~\cite{Dubey2021}. 
When $\tau > \tau_Z$, Eq.~\eqref{eq:mapping} does no longer hold and, in the time interval between consecutive measurements, the wavefunction has sufficient time to spread significantly into the monitored region.
In this regime, we solve the problem numerically: the system is evolved in time using a split-step method~\cite{Taha1984} and, after each interval $\tau$, we compute $F_n$ as in Eq.~\eqref{eq:norm}.
The wavefunction is then set to zero within the detection region~\cite{supplementary}, and the time evolution is continued accordingly.
\begin{figure}
    \centering
    \includegraphics[width=0.9\linewidth]{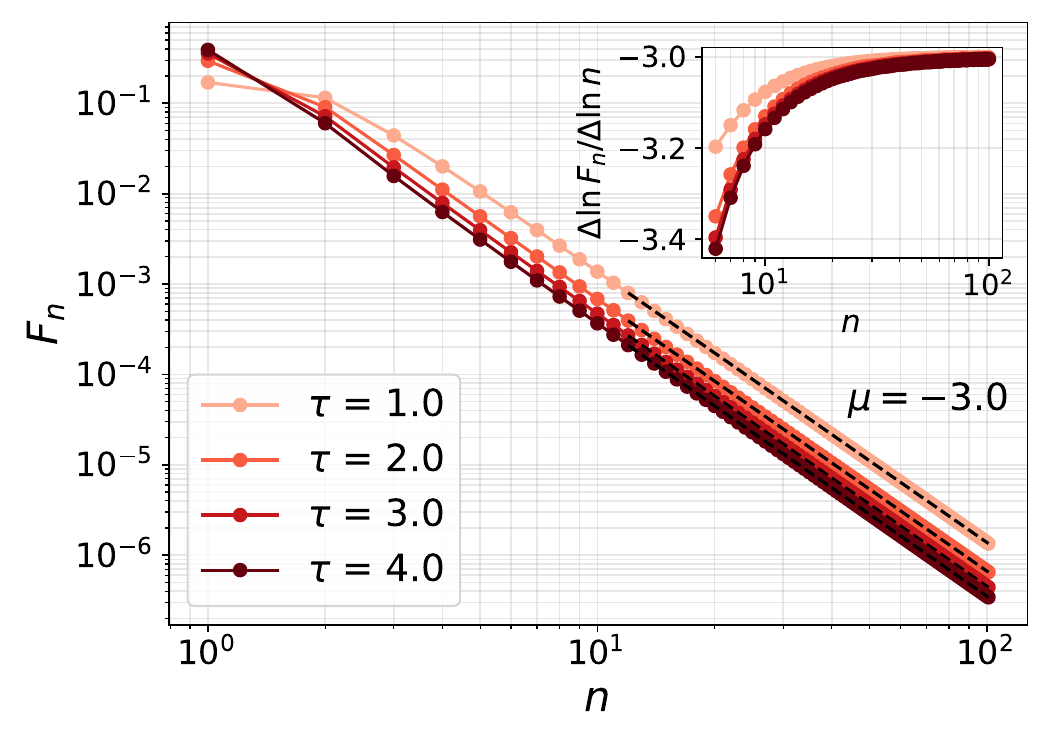}
    \caption{Probability $F_n$ of first detection in $x>0$ at the $n$-th measurement for a free particle with $x_0 = -1$, $\sigma=0.8$, and for various values of the measurement interval $\tau$ (ranging from 1 to 4)~\cite{femto}. The data clearly show an algebraic decay as a function of $n$: The dashed black line correspond to the numerical fit  $F_n \propto n^{\mu}$, yielding $\mu \simeq -3$. The inset shows the effective exponent of the algebraic dependence of $F_n$ on $n$ for $n \gtrsim 6$, showing convergence towards $-3$.
    This behavior emerges independently of the value of $\tau$ and of the choice of $x_0$ and $\sigma$.
    }
    \label{fig:Fn_nz}
\end{figure}
Figure~\ref{fig:Fn_nz} displays the  dependence of $F_n$ on $n$ for various values of $\tau$, clearly showing an algebraic decay for $n \gg 1$, consistent with the one $F_n \propto n^{-3}$  we found in the Zeno regime.

\begin{figure}[h]
    \centering
    \includegraphics[width=1\linewidth]{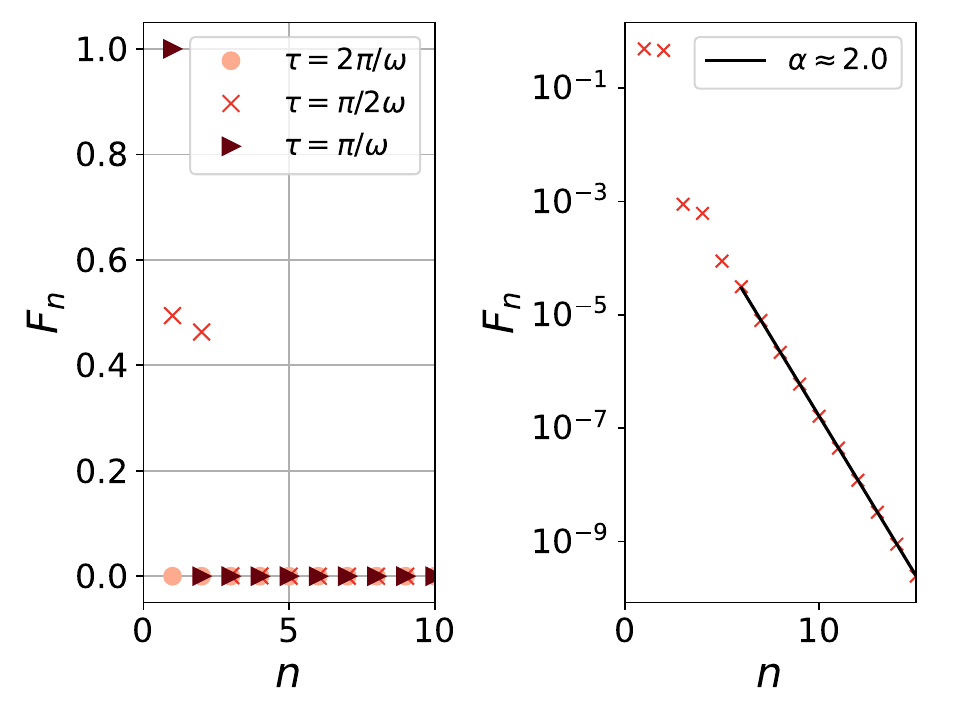}
    \caption{First-detection probability $F_n$ for the harmonic oscillator as a function of $n$. In the left panel $F_n$ is shown for $\omega = 2.45$ and various (possibly resonant) duration of the interval $\tau$ between successive  measurements. For resonant values of $\tau$, $F_n$ may either vanish identically when measured at multiples of the period (as in the case $\tau=2\pi/\omega$ here) or it vanishes after the first  measurement when measured at half the period (case $\tau=\pi/\omega$). Otherwise, $F_n$ decays exponentially to zero upon increasing $n$ (case $\tau=\pi/(2 \omega)$), as highlighted in the log-linear plot on the right, with $F_n \propto e^{-\alpha n \tau}$ and $\alpha\simeq 2.0$
    }
    \label{fig:comp_sca_ho}
\end{figure}

\emph{Harmonic oscillator.---} In contrast with the case of a delocalized free particle, we now consider a particle fully confined by an harmonic potential centered at the origin, with Hamiltonian
\begin{equation}
    H = p^2/2+ \omega^2\, x^2/2 \label{eq:Hamiltonian_HO}.
\end{equation}
When $\tau \ll \tau_Z$ [see Eq.~\eqref{eq:def-tauZ}], the mapping in Eq.~\eqref{eq:mapping} for the free particle can be applied by extending  
the construction reported in Refs.~\cite{Allcokc1969-2, Halliwell2010}. 
Moreover, since the condition which determines the value of $V_0$ in Eq.~\eqref{eq:mapping}
depends on the commutator $[H,D]$ between the Hamiltonian $H$ and the projector $D = \theta(-x)$, adding a potential $U(x) = \omega^2 \, x^2/2$ that commutes with $\theta(-x)$ does not modify the conditions under which Eq.~\eqref{eq:mapping} applies. 
In this regime, the time-independent Schr\"odinger equation takes the form
\begin{equation}
    -\frac{1}{2}\xi_l^{\prime\prime}(x) +\frac{1}{2}\omega^2x^2\xi_l(x) - iV_0\theta(x)\xi_l(x) = E_l\xi_l(x),
\end{equation}
where $\xi_l(x)$ and $E_l$ are the eigenfunctions and the eigenvalues, respectively, of the complex effective Hamiltonian $H_{\rm eff} \equiv H-iV_0\theta(x)$. 
The general solution of this equation can be written in terms of the parabolic cylinder functions $D_\nu(z)$ as~\cite{abramowitz}
\begin{equation}
    \xi_l(x) = \begin{cases}
        A_{-,l} D_{\nu_{-,l}} (2\sqrt{\omega}|x|) &\quad\mbox{for}\quad  x<0,\\
        A_{+,l} D_{\nu_{+,l}} (2\sqrt{\omega}x)&\quad\mbox{for}\quad  x>0,
    \end{cases}
\end{equation}
with amplitudes $A_{\pm,l}$ and indexes $\nu_{\pm,l}$ to be determined. 
The continuity of $\xi_l(x)$ and $\xi_l'(x)$ at $x=0$ eventually yields the eigenvalue equation
\begin{equation}
    \frac{\Gamma(\frac{1-\nu_{+,l}}{2})}{\Gamma(\frac{-\nu_{+,l}}{2})} +\frac{\Gamma(\frac{1-\nu_{-,l}}{2})}{\Gamma(\frac{-\nu_{-,l}}{2})} = 0,
\end{equation}
where $\nu_{-,l} = E_l/\omega -1/2$ and $\nu_{+,l} = (E_l+iV_0)/\omega -1/2$. 
This implicit equation is solved by a set of complex values $E_l = \Omega_l - i \gamma_l$ with real and imaginary parts $\Omega_l$ and $\gamma_l$, respectively. 
Then, the time-dependent wavefunction at the stroboscopic times $n\tau$ 
can be written as
\begin{equation}
    \psi(x,n\tau) = \sum_l c_l\xi_l(x)e^{-iE_l n\tau},
    \label{eq:time_evo_ho}
\end{equation}
where $c_l$ denotes the overlap between the initial condition $ \psi(x,t=0)$ and the left eigenfunction $\tilde{\xi_l}$ of $H_{\rm eff}$ with eigenvalue $E_l$.
The presence of a nonzero imaginary part $\gamma_l$ in $E_l$ causes each mode to decay exponentially upon increasing time $n$, leading to an overall exponential suppression of the detection probabilities $F_n$ (see Eq.~\eqref{eq:norm}). 
In contrast with the case of a free particle, where the continuous spectrum yields long-time algebraic tails, the discrete nature of the spectrum of the harmonic oscillator suppresses such behavior, resulting in this exponential decay.
Indeed~\cite{supplementary}, by inserting Eq.~\eqref{eq:time_evo_ho} into Eq.~\eqref{eq:norm}, one finds
\begin{equation}
    F_n  = \sum_{m,k} c_mc^*_kR_{m,k}e^{-i(\Omega_m - \Omega_k)n\tau}e^{-\left(\gamma_m+\gamma_k\right)n\tau}, \label{eq:Fn_h}
\end{equation}
where $R_{m,k} = \int_0^\infty \dd x\,\xi_m(x)\xi_k^*(x)$.
For small $\tau$, i.e., for $V_0 \gg 1$ and for $n \gg 1$, the most relevant contribution to Eq.~\eqref{eq:Fn_h} is given by the term with the smallest $\gamma_n$, i.e.,
\begin{equation}
    F_n \propto e^{-2\tilde{\gamma}n\tau}, \quad \mbox{where}\quad \tilde{\gamma} = \min_{n}\gamma_n.
    \label{eq:exp-decay}
\end{equation}
As previously discussed, for arbitrarily large $\tau$, the representation of the measurement in terms of an  
imaginary potential is no longer applicable and thus one has to resort to numerical analysis. 
Contrary to the case of a free particle, the confining nature of the potential gives rise 
to an interesting behavior of $F_n$. 
Let us show that by analyzing the case of a weakly confining harmonic potential, i.e., $\omega \lesssim (\sigma^2\tau)^{-1/2}$, 
with the particle that is initially localized close to the minimum of this potential. 
\begin{figure}
    \centering
    \includegraphics[width=0.9\linewidth]{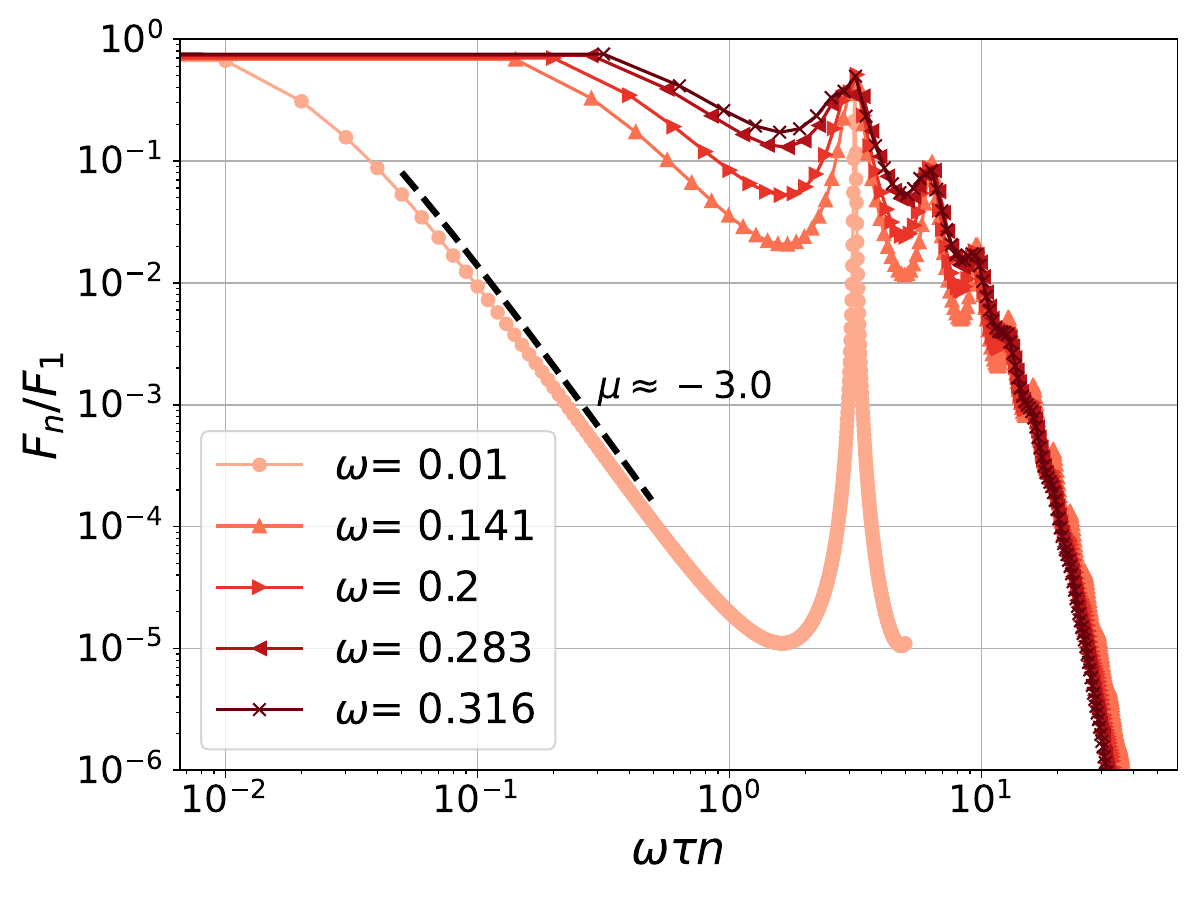}
    \caption{First-detection probability $F_n$ for the harmonic oscillator, with initial condition given in Eq.~\eqref{eq:iintial_free} and parameters $x_0 = -1$, $\sigma = 0.8$, as a function of $n$ on a log-log scale and for various values of the harmonic potential strength $\omega^2$. A behavior $F_n \propto n^{-3}$ consistent with that of the free particle is observed at small values of $\omega \lesssim (\sigma^2\tau)^{-1/2}$,  as highlighted by the dashed line; for values of $n\omega\tau = \kappa\pi$ with integer $\kappa =1, 2, \dots$, resonance due to the harmonic confinement emerge.} 
    \label{fig:F_n_sca_sma_k}
\end{figure}
%
%
As expected, upon increasing $n$, in Fig.~\ref{fig:F_n_sca_sma_k} we observe an initial algebraic decay of $F_n$, which is similar to that observed for a free particle.
At larger $n$, however, the confining potential starts being seen, resulting into a series of resonances in $F_n$, characterized by peaks, followed by an exponential decay of $F_n$~\footnote{The same analysis with $\omega$ fixed and varying $\tau$ reproduces similar results, with resonances at multiples of $\omega\tau$.}. The occurrence of resonances is due to the confining nature of the potential. Indeed, some simple resonances can be understood from the motion of the center $x_c(t)$ of a Gaussian wave packet, initially at $x_c(0)=x_0<0$, in a harmonic potential follows the classical trajectory
\begin{equation}
    x_c(t) = x_0\cos(\omega t). \label{eq:mo_Ho}
\end{equation}
This implies that the particle is entirely localized in the monitored region $x>0$ at the first measurement whenever the sampling time $\tau$ is an odd (integer) multiple of $\pi/\omega$, yielding a detection probability $F_1 = 1$.
Conversely, if $\tau$ is an even multiple of $\pi/\omega$ 
and assuming that $|x_0| \gtrsim 3\sigma$~\footnote{When this condition is fulfilled, the Gaussian tail of the initial wavepacket yields a practically negligible detection probability.}, 
the particle will never be within the detection region at the time of measurements and thus it remains practically undetected.
This behavior is illustrated in Fig.~\ref{fig:comp_sca_ho}, where, on the left, $F_n$ is plotted as a 
function of $n$ for various values of $\tau$, either resonant or not.  In particular, in the latter case, 
a generic exponential decay of $F_n$ upon increasing $n$ is observed, as highlighted  in the right panel of the figure.
\emph{Potential well.---} We now turn to a case which is intermediate between those discussed above, namely a particle in a potential well $U(x)$ of finite depth $U_0 < 0$ and width $a$, i.e.,
\begin{equation}
    U(x) = \left\{\begin{split}
            U_0 &\quad\mbox{for}\quad |x|<a,\\
            0  &\quad\mbox{for}\quad |x|\ge a.
        \end{split}\right.
\end{equation}
This potential supports both bound states (i.e., a discrete spectrum) and a continuum of scattering states, thus combining the features of the harmonic confinement and of the free particle.
Depending on the initial state, this mixed spectrum may result in a first-detection probability $F_n$ which decreases upon increasing $n$ either algebraically or exponentially.
As in the previous cases, in the Zeno limit one can exploit the mapping in Eq.~\eqref{eq:mapping}, allowing an analytical treatment of the dynamics.
In fact, the wave function in the detection region can be written as~\cite{supplementary}
\begin{equation}
\begin{split}
    \psi(x,n\tau^-) &= \sum_k c_k \phi_k(x)e^{-iE_kn\tau^-} \\
    &\quad + \int \!\dd k \,b(k)\tilde{t}(k) e^{-iq(k)n\tau^-}e^{-iE(k)n\tau^-},
\end{split}
    \label{eq:well}
\end{equation}
where the bound states $\{\phi_k(x)\}$ arise from the poles of the scattering matrix, while the scattering states $\propto \tilde{t}(k) e^{-iq(k)t}$ on the second line of the equation emerge from the continuum part of the spectrum.
The first term on the r.h.s.~of Eq.~\eqref{eq:well} leads to an exponential decay of $F_n$, originating from the nonzero imaginary parts of the quasi-bound 
state energies $\{E_k\}$ (due to ${\rm Im}\, E_k \neq 0$, these states vanish under time evolution). 
In contrast, the second term, associated with the continuum part of the spectrum, yields an algebraic decay analogously to the case of a free particle discussed above.
The relative weight of these two contributions can be controlled by an appropriate choice of the initial state, which allows one to control the asymptotic behavior of $F_n$.
The interplay between the exponential and the algebraic decay persists even for larger values of $\tau$, where the mapping in Eq.~\eqref{eq:mapping} no longer applies. In this system, the fundamental distinction between a classical first-passage problem and its quantum counterpart becomes particularly evident. In fact, the quantum measurement process is intrinsically active, capable of both 
injecting and removing energy from the system through two competing 
mechanisms. On the one hand, localizing the particle on one side of the well 
after an unsuccessful detection effectively injects energy into the system.
On the other, the projection removes the high-energy components of the wavefunction that 
first reach the measured region, thereby reducing the system's energy.
At the beginning of the dynamics, when the confining effect of the measurement is dominant, this energy injection can facilitate the escape of the particle  from the potential well. Consequently, one expects a crossover in the dependence of $F_n$ on $n$ from an exponential to an algebraic behavior upon increasing $n$.
\begin{figure}[t]
    \centering
    \includegraphics[width=1\linewidth]{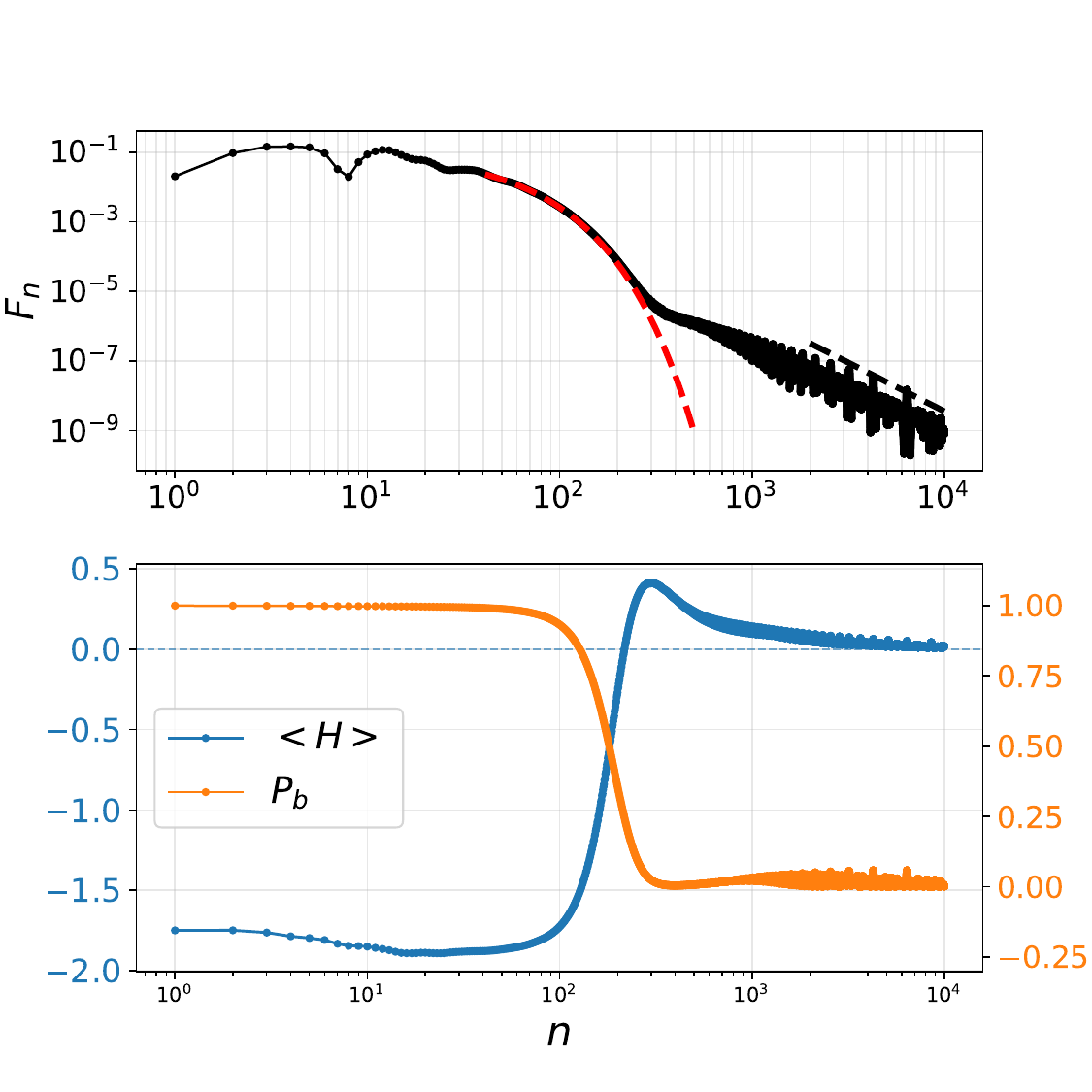}
    \caption{Dependence on $n$ of the first-detection probability $F_n$ (upper panel), of the average 
energy $\langle H \rangle$ (lower panel, blue) and of the probability $P_b$ (lower panel, orange) of being in a bound state, for a particle moving in a potential well of finite depth $U_0 < 0$.  The upper panel shows that an initial exponential decay (red dashed line) of $F_n$ at intermediate values of $n$ is followed by an algebraic decay $F_n\propto n^{\mu}$ with $\mu \propto 2.8$. 
The crossover between these two behaviors occurs when the average 
energy (lower panel) becomes positive and the delocalized states become more relevant in the particle’s wave function, as confirmed by the fact that probability $P_b$ of being in a bound state approaches zero. Simulations correspond to $a=5.0$, $U_0 = -2$, $x_0 = -2$, and $\sigma = 1$.}
    \label{fig:energy}
\end{figure}
%
This is demonstrated in Fig.~\ref{fig:energy}, where the dependence of $F_n$ on $n$ (upper panel) is compared with that of the mean energy $\langle H \rangle$ (lower panel), which is constant between two consecutive measurements. As long as the particle is likely to remain trapped in the potential well, so that $\langle H \rangle < 0$, the decay of $F_n$ as a function of $n$ is exponential. Once the 
particle acquires a positive energy, i.e., $\langle H \rangle > 0$, indicating that it has escaped 
the well, the behavior crosses over to an algebraic decay. This kind of cross over seems to be a general, peculiar, and interesting property of the statistics of certain first-detection problems, and not specific to our model. Indeed, this is suggested both by the rather general mechanism described here and by the presence of a similar cross over in a many-body first-detection problem~\cite{Walter_2025}.

\emph{Conclusions.---} In summary, we investigated the statistics of the first-detection time for a quantum particle in one spatial dimension, subject to stroboscopic projective measurements. 
By analyzing the cases of a free particle, of the (confined) harmonic oscillator, and the intermediate case of a potential well of finite depth, we show that a connection exists between the behavior of the probability $F_n$ for the first detection to occur at the $n$-th measurement and the nature of the spectrum of the underlying Hamiltonian. 
In the Zeno regime, the dynamics can be mapped onto an effective evolution under an imaginary potential, see Eq.~\eqref{eq:mapping}, providing analytical access to the behavior of $F_n$, while beyond this regime, numerical simulations reveal a distinct crossover from an exponential to an algebraic decay. 
Our results highlight the intrinsically active role of measurement in quantum first-passage processes: measurement back action not only perturbs but fundamentally reshapes the distribution of the energy of  the  system and escape dynamics. 
The interplay among measurement, confinement, and spectral structure paves the way for controlled experimental realizations in engineered quantum systems.

\emph{Acknowledgement---}
GDF thanks E.~Barkai for stimulating comments and insights.  
GDF and AC thank A.~Scardicchio and J.~Niedda for valuable discussion. 
GDF, AG, and AS acknowledge support from the PNRR MUR project PE0000023--NQSTI.

\bibliographystyle{apsrev4-2}

\begin{thebibliography}{35}
\bibitem [{fem()}]{femto}%
	\BibitemOpen
	\href@noop {} {\bibinfo {title} {{If we choose the unit of length as $ L_0 =
				1\,\text{nm}$, then $\tau = 1 $ in our rescaled units corresponds to
				approximately 10 femtoseconds.}}}\BibitemShut {Stop}%
\end{thebibliography}

\end{document}


\clearpage

\widetext
\pagebreak
\setcounter{equation}{0}
\setcounter{figure}{0}
\setcounter{table}{0}
\setcounter{page}{1}
\renewcommand{\theequation}{S\arabic{equation}}
\setcounter{figure}{0}
\renewcommand{\thefigure}{S\arabic{figure}}
\renewcommand{\thepage}{S\arabic{page}}
\renewcommand{\thesection}{S\arabic{section}}
\renewcommand{\thetable}{S\arabic{table}}
\makeatletter

\renewcommand{\thesection}{\arabic{section}}
\renewcommand{\thesubsection}{\thesection.\arabic{subsection}}
\renewcommand{\thesubsubsection}{\thesubsection.\arabic{subsubsection}}

\renewcommand{\bibnumfmt}[1]{[S#1]}

\renewcommand{\citenumfont}[1]{S#1}
\newcommand{\aldo}[1]{\textcolor{orange}{#1}}

	\newpage
	\onecolumngrid
	\bigskip 
	
	\begin{center}
		\large{\bf Supplemental Material to "Measurement-induced crossover in quantum first-detection times" \\}
    \author{Giovanni Di Fresco}
    \affiliation{SISSA-International School for Advanced Studies and INFN, via Bonomea 265, 34136 Trieste, Italy}
    \author{Aldo Coraggio}
    \affiliation{SISSA-International School for Advanced Studies and INFN, via Bonomea 265, 34136 Trieste, Italy}
    \author{Alessandro Silva}
    \affiliation{SISSA-International School for Advanced Studies, via Bonomea 265, 34136 Trieste, Italy}
    \author{Andrea Gambassi}
    \affiliation{SISSA-International School for Advanced Studies and INFN, via Bonomea 265, 34136 Trieste, Italy}

	\end{center}

\vspace{1cm}
This Supplemental Material is organized in the following sections:
\begin{enumerate}
\item Regularization of projective measurements and mapping to non-Hermitian dynamics,
\item $F_n$ for the free particle in the Zeno regime,
\item First detection probability for non-Hermitian harmonic oscillator,
\item First detection probability for a potential well,
\item First-passage-time statistics of the Wiener process subject to stroboscopic measurements.
\end{enumerate}

\section*{1. Regularization of Projective Measurements and Mapping to Non-Hermitian Dynamics}

\subsection*{1A. Sharp position measurements and their pathologies}

In the main text of the present work, we are interested in detecting whether a quantum particle is located on one side of a sharp spatial boundary, e.g., on $x \ge 0$, at every time interval $\tau$, computing probability $F_n$ that this detection occurs for the first time at the $n$-th measurement. After an unsuccessful detection, the state of the particle before the detection is projected onto the no-detection outcome which, in quantum mechanics, is represented by the projector
\begin{equation}
    P_{x<0} = \theta(-x).
\label{eqS:defP}
\end{equation}
where $\theta(x)$ denotes the Heaviside step function. When this projector acts on the pre-measurement wavefunction $\psi_0(x)$, the post-measurement state  $\psi(x)$ becomes
\begin{equation}
    \psi(x) = \frac{\theta(-x)\psi_0(x)}{\|\theta\psi_0\|},
\end{equation}
where $\|\theta\psi_0\|$ is the norm of  $\theta(-x)\psi_0(x)$. Note that $\psi(x)$ may be discontinuous at $x=0$ even if $\psi_0(x)$ was not. The Fourier transform $\widetilde\psi(p)$ of $\psi(x)$ therefore develops algebraic tails, i.e., $\widetilde\psi(p)\!\sim\!1/p$, implying a divergent momentum variance and hence an infinite kinetic energy in the state $\psi$, i.e., 
\begin{equation}
    \langle T\rangle = \frac{\langle p^2\rangle}{2m} = \infty.
\end{equation}
This divergence actually reflects the Heisenberg uncertainty principle
\begin{equation}
    \Delta x\,\Delta p \gtrsim \frac{\hbar}{2}.
\end{equation}
In fact, a perfectly sharp boundary of the detection region corresponds to having a vanishing uncertainty $\Delta x$ in the determination of the position of the particle, i.e., $\Delta x \to 0$, which implies $\Delta p\to\infty$ and thus makes $\langle p^2\rangle$ diverge. 
Accordingly, although mathematically consistent, infinitely sharp projectors are physically ill-defined, since no realistic detector can perform measurements with infinite spatial resolution. In the next two subsections, we show how the position measurement introduced in the main text can be regularized.

\subsection*{1B. Regularization in the small-$\tau$ regime}\label{sec:map}

When projective measurements are repeated at short time intervals $\tau$, the system enters the quantum Zeno regime. The propagator can be expressed as a string of projectors,
\begin{equation}
    e^{-iH\tau/\hbar}\left(P_{x<0}\,e^{-iH\tau/\hbar}\right)^{n-1}, 
    \label{eqS:decompos}
\end{equation}
where $P_{x<0}$ is the projector defined in Eq.~\eqref{eqS:defP} which, as discussed above, leads to discontinuous post-measurement states.  
Following Refs.~\cite{Allcokc1969-2, Halliwell2010}, this pulsed-measurement sequence can be approximated by a continuous evolution under a non-Hermitian Hamiltonian,
\begin{equation}
    H_{\mathrm{eff}} = H - iV_0\,\theta(x),
    \label{eqS:Heff}
\end{equation}
where $V_0$ represents an absorbing potential.  
The correspondence between the pulsed and continuous evolution can roughly be understood by noting that
\begin{equation}
    e^{-V_0\theta(x)\tau} = \theta(-x) + e^{-V_0\tau}\left[1- \theta(-x)\right].    
\end{equation}
Accordingly, as long as $e^{-V_0\tau}\ll 1$,
\begin{equation}
    \theta(-x)\simeq e^{-V_0\theta(x)\tau},\label{eq:map}
\end{equation}
and thus, form Eq.~\eqref{eqS:defP}, one can approximate $P_{x<0} \simeq e^{-V_0\theta(x)\tau}$ in Eq.~\eqref{eqS:decompos}, from which one easily recognizes the evolution under $H_{\rm eff}$ in Eq.~\eqref{eqS:Heff}.
We refer to Ref.~\cite{Halliwell2010} for a more detailed and rigorous derivation, which shows that Eq.~\eqref{eq:map} is a good approximation for describing the dynamics as long as 
\begin{equation}
    V_0\tau^2 |\langle[H, \theta(x)]\rangle|\ll 1.
    \label{eqS:V0tau}
\end{equation}
In the case of a free particle, this holds for
\begin{equation}
    V_0 \tau \approx 4/3.
     \label{eqS:V0tau-fp}
\end{equation}
Indeed, frequent measurements ($\tau \to 0$) correspond to a strong absorber ($V_0 \to \infty$).
Unlike the sharp projection, a finite $V_0$ does not truncate the wavefunction abruptly. Instead, it causes the wavefunction that ``leaks'' into $x>0$ to decay exponentially over a length scale
\begin{equation}
    \ell = \hbar/\sqrt{mV_0},
\end{equation}
producing a continuous wavefunction at the boundary of the detection region. 
The wavefunction evolved with $H_{\rm eff}$ [see Eq.~\eqref{eqS:Heff}] up to time $n\tau$ corresponds, within the mapping errors, to the wavefunction of the actual system subject to stroboscopic measurements just before the $n$-th one, allowing us to work with a well-behaved wavefunction and a regularized dynamics that avoids divergent kinetic energies.

\subsection*{1C. Large-$\tau$ regime and general regularization: finite-resolution measurements}
The mapping of the measurement process onto the evolution with an imaginary potential is accurate for sufficiently small $\tau$, but some form of regularization must also be adopted for arbitrary intervals between successive measurements.
A general and physically motivated approach is to assume that the detector of the particle position has a finite spatial resolution, which practically smooths the step function $\theta(-x)$ present in $P_{x<0}$ and yields the measurement operator
\begin{equation}
    \chi_\varepsilon(x) = \frac{1+\operatorname{erf}\!\left(x/\varepsilon\right)}{2},
\end{equation}
where $\varepsilon$ characterizes the boundary width and $\erf$ indicates the error function. Accordingly, a measurement on the state described by $\psi(x)$ produces the post-measurement state
\begin{equation}
    \psi_\varepsilon(x) = \frac{\chi_\varepsilon(x)\,\psi(x)}{\|\chi_\varepsilon\psi\|},
\end{equation}
which is smooth across $x=0$ and has finite kinetic energy 
\begin{equation}
    \langle T\rangle \simeq \frac{\hbar^2}{2m}\,\frac{C}{\varepsilon^2},
\end{equation}
where $C$ is a constant depending on $\psi(x)$.
The operator $\chi_\varepsilon(x)$ represents a position measurement with finite spatial resolution $\Delta x \sim \varepsilon$, consistent with the uncertainty principle and with realistic detector resolution. Note, however, that $\chi_\varepsilon(x)$ is not a projector; nonetheless, it can be realized as a positive operator-valued measure (POVM)~\cite{Jacobs2014}, obtained from a standard projective measurement on an enlarged Hilbert space.
In particular, this can be done considering a two-level ancilla with Kraus operators
\begin{equation}
    A = \chi_\varepsilon(\hat X), \qquad 
    B = \sqrt{\mathbb I - \chi^2_\varepsilon(\hat X)},
\end{equation}
which satisfy $A^2 + B^2 = \mathbb I$. The system–ancilla state prior to the measurement is initialized as
\begin{equation}
    \ket{\Psi_{\text{in}}} = \int \!\dd x \, \psi(x)\ket{x} \otimes \ket{0}.
\end{equation}
The unitary interaction
\begin{equation}
    U = 
    \begin{pmatrix}
        A & -B\\[4pt]
        B & A
    \end{pmatrix}
\end{equation}
acting on the composite system of the particle and the ancilla leads to the final state
\begin{equation}
    \ket{\Psi_{\text{out}}} =
    \left(\int\! \dd x \, A\,\psi(x)\ket{x}\right) \otimes \ket{0}
    +
    \left(\int\! \dd x \, B\,\psi(x)\ket{x}\right) \otimes \ket{1}.
\end{equation}
Performing a projective measurement on the ancilla and post-selecting the outcome $\ket{0}$ yields the desired state. Physically, this procedure corresponds to a position measurement with finite spatial resolution implemented via a probe system.

Sharp position projections lead to unphysical divergences in relevant mechanical quantities, such as the kinetic energy, due to discontinuities in the projected wave function at the measurement boundary. Regularizing the measurement process, either by mapping it to an absorbing potential (in the Zeno limit of small $\tau$) or by explicitly smoothing the spatial cutoff (for arbitrary $\tau$), eliminates these divergences while preserving the essential physics of detection. Both approaches admit a clear operational interpretation: realistic position measurements possess an intrinsically finite spatial resolution, and their repeated application can be modeled consistently without violating the Heisenberg uncertainty principle.

\section{$F_n$ for the Free Particle in the Zeno regime}

When $\tau < \tau_Z$, as defined in Eq.~(7), we can exploit the mapping to a non-Hermitian potential discussed in Eq.~(8) of the main text and in the previous section  in order to derive an analytical expression for $F_n$ and to determine its dependence on $n$. The dynamics is given by the effective Hamiltonian $H_{\text{eff}}$ in Eq.~\eqref{eqS:Heff}, which takes the form of a scattering problem from a constant non-Hermitian potential. The wavefunction can therefore be expressed as 
\begin{equation}
    \psi(x, t) = \left\{\begin{split}
        \psi_{I} + \psi_{R} &\quad \mbox{for}\quad x<0, \\
        \psi_{T} &\quad \mbox{for}\quad x>0.\\
    \end{split}\right.
\end{equation}
Accordingly, the scattering state can be written as 
\begin{equation}
   \phi_k(x) = \left\{\begin{split}
        e^{ikx} + R(k)e^{-ikx} &\quad \mbox{for}\quad x<0, \\
        T(k)e^{iq(k)x} &\quad \mbox{for}\quad  x>0, \\
    \end{split}\right.
    \label{eqS:phik}
\end{equation}
where the reflection and transmission coefficients are, respectively, given by
\begin{equation}
    R(k)= \frac{k-q(k)}{k+q(k)}\quad\mbox{and}\quad T(k) = \frac{2k}{k+q(k)} \quad\mbox{where} \quad q(k) = \sqrt{k^2 + 2iV_0}.
    \label{eq:ref_tran}
\end{equation}
For the square root in the latter expression, we choose the branch of the square root which has $\Im(q(k))>0$. The time-dependent wavefunction thus reads
\begin{equation}
    \psi(x,n\tau) = \left\{\begin{split}
                \int\!\frac{\dd k}{\sqrt{2\pi}}\widetilde{\psi}(k)\left[ e^{ikx} + R(k)e^{-ikx}\right]e^{-ik^2 n\tau/2} &\quad \mbox{for}\quad  x<0,\\
                 \int\!\frac{\dd k}{\sqrt{2\pi}}\widetilde{\psi}(k)T(k)e^{-iq(k)x}e^{-ik^2n\tau/2} &\quad \mbox{for}\quad  x>0,
    \end{split}\right.\label{eq:wave_function_time}
\end{equation}
where $\widetilde{\psi}(k)$ denotes the Fourier transform of the initial state.
In particular, for $V_0 \gg 1/(4\sigma^2)$, one may expand Eq.~\eqref{eq:ref_tran} as
\begin{equation}
        q(k) \simeq \sqrt{2iV_0}\left[1 + \frac{k^2}{4iV_0}  + O \left(k^4/V_0^2\right)\right]\quad \mbox{and} \quad T(k) \simeq \frac{2k}{\sqrt{2iV_0}},
\end{equation}
while the transmitted wavefunction takes the form
\begin{equation}
    \psi(x,n\tau) = \mathcal{A}e^{i\sqrt{2iV_0}x}\int\!\! \dd k \, k\, e^{-\sigma^2k^2-ikx_0} e^{ik^2/\sqrt{8iV_0}}e^{-ik^2n\tau/2},
\end{equation}
where $\mathcal{A} = \frac{2^{3/4}(\pi\sigma^2)^{1/4}}{\sqrt{2\pi^2iV_0}}$. 
By calculating this Gaussian integral, we arrive at
\begin{equation}
    |\psi(x,n\tau)|^2 =|\mathcal{A}|^2 \left|\frac{B}{2A} \sqrt{\frac{\pi}{A}} e^{B^2/(4A)}\right|^2 e^{-2\sqrt{V_0}x},\label{eq:psi2}
\end{equation}
where $A = \sigma^2 + \frac{i n\tau}{2} - \frac{i x}{\sqrt{8 i V_0}}$ and $B = i x_0$. According to Eq.~(4) of the main text, $F_n$ is given by $F_n = \int_0^{\infty}\dd x\, |\psi(x,n\tau)|^2$.
From Eq.~\eqref{eq:psi2} it is clear that $|\psi(x,n\tau)|^2$ decays exponentially upon increasing $x$ and becomes negligible for $x > 1/(2\sqrt{V_0})$. Consequently, the $x$-dependence of $A$ can be neglected, while its dependence on time introduces the dependence on $n$. This yields an expression for $F_n$ of the form
\begin{equation}
F_n = |\mathcal{A}|^2 
\left|\frac{B}{2A} \sqrt{\frac{\pi}{A}} e^{B^2/(4A)}\right|^2 
\frac{1}{2\sqrt{V_0}}, 
\qquad \text{with} \qquad 
A \simeq \sigma^2 + \frac{i n\tau}{2}.
\label{eq:appA_sca}
\end{equation}
Since the mapping discussed in the previous section implies $V_0 \simeq 1/\tau$ [see Eqs.~\eqref{eqS:V0tau} and, for the free particle, \eqref{eqS:V0tau-fp}], Eq.~\eqref{eq:appA_sca} yields, for sufficiently large $n$,
\begin{equation}
    F_n \simeq \tau^{3/2} n^{-3},
\end{equation}
which proves Eq.~(12) of the main text.

\section*{2. First Detection probability for Non-Hermitian Harmonic Oscillator}
By exploiting the mapping discussed in the main text and in the previous sections, the problem of the first-detection time for the harmonic oscillator in the Zeno regime can be conveniently studied in terms of a non-Hermitian Hamiltonian.
\begin{equation}
H = \frac{p^2}{2} + \frac{1}{2} \omega^2 x^2 \;-\; i V_0 \,\theta(x), \quad\mbox{with} \quad   V_0>0. \label{eqS:H}
\end{equation}
In this section we (i) derive the stationary modes of $H$ and the corresponding energy quantization condition;  
(ii) construct the time-dependent mode expansion of the wavefunction, and  
(iii) determine the asymptotic decay of the first-detection probability $F_n$.

The time-independent Schr\"odinger equation associated to the (non-Hermitean) Hamiltonian in Eq.~\eqref{eqS:H} reads
\begin{equation}
-\frac{1}{2}\,\xi_l''(x) + \frac{1}{2}\,\omega^2 x^2\,\xi_l(x) - i V_0\,\theta(x)\,\xi_l(x) = E_l\,\xi_l(x),
\label{eq:TISE}
\end{equation}
where $\xi_l(x)$ is the right eigenfunction of $H$ with eigenvalue $E_l$.
Introducing $z=\sqrt{2\omega}\,x$ and
\begin{equation}
\nu_{-, l}=\frac{E_l}{\omega}-\frac{1}{2},\qquad
\nu_{+, l}=\frac{E_l+iV_0}{\omega}-\frac{1}{2} = \nu_{-,l} + i \frac{V_0}{\omega},
\label{eq:nus}
\end{equation}
Eq.~\eqref{eq:TISE} reduces, both for $x>0$ and $x<0$, to the Weber equation~\cite{abramowitz}, the solution of which are parabolic-cylinder functions $D_\nu(z)$. 
The solutions which decay to zero as $|x| \to \infty$  are
\begin{equation}
    \xi_l(x) = \begin{cases}
        A_{-,l} D_{\nu_{-,l}} (2\sqrt{\omega}|x|) &\quad\mbox{for}\quad  x<0,\\
        A_{+,l} D_{\nu_{+,l}} (2\sqrt{\omega}x)&\quad\mbox{for}\quad  x>0.
    \end{cases}
    \label{eqS:xi}
\end{equation}
This solution vanishes for $|x|\to \infty$ because $D_\nu(z)\sim z^\nu e^{-z^2/4}$ as $z\to+\infty$~\cite{abramowitz}. Requiring the continuity of $\xi_l(x)$ and $\xi_l'(x)$ at $x=0$, with  $\xi_l(0^\pm)=A_{\pm, l} D_{\nu_{\pm, l}}(0)$ and $\xi_l'(0^\pm)=\pm A_{\pm,l}\sqrt{2\omega}\,D'_{\nu_{\pm,l}}(0)$, 
yields the conditions
\begin{equation}
\begin{pmatrix}
D_{\nu_{-,l}}(0) & -D_{\nu_{+,l}}(0)\\[2pt]
-\sqrt{2\omega}\,D'_{\nu_{-,l}}(0) & -\sqrt{2\omega}\,D'_{\nu_{+,l}}(0)
\end{pmatrix}
\begin{pmatrix} A_{-,l}\\ A_{+,l}\end{pmatrix}=\bm{0},
\end{equation}
which admits nontrivial solutions only if the determinant of the matrix on the l.h.s.~of the equation above vanishes, i.e., if
\begin{equation}
D_{\nu_{-,l}}(0)\,D'_{\nu_{+,l}}(0)+D_{\nu_{+,l}}(0)\,D'_{\nu_{-,l}}(0)=0.
\label{eq:det0}
\end{equation}
Dividing this condition by $D_{\nu_{+,l}}(0)D_{\nu_{-,l}}(0)$ gives
\begin{equation}
-\,\frac{D'_{\nu_{-,l}}(0)}{D_{\nu_{-,l}}(0)} \;=\; \frac{D'_{\nu_{+,l}}(0)}{D_{\nu_{+,l}}(0)}.
\label{eq:ratio}
\end{equation}
Using the known values of $D_\nu$ and $D'_\nu$ at the origin, i.e.,~\cite{abramowitz}
\begin{equation}
D_{\nu}(0)=\frac{2^{\nu/2}\sqrt{\pi}}{\Gamma\!\left(\frac{1-\nu}{2}\right)},\quad\mbox{and}\quad
D'_{\nu}(0)=-\frac{2^{(\nu+1)/2}\sqrt{\pi}}{\Gamma\!\left(-\frac{\nu}{2}\right)},
\label{eq:D0}
\end{equation}
we arrive at the transcendental quantization condition for $\nu_{\pm}$ given in Eq.~\eqref{eq:nus}:
\begin{equation}
\frac{\Gamma\big(\tfrac{1-\nu_{+,l}}{2}\big)}{\Gamma\big(-\tfrac{\nu_{+,l}}{2}\big)} +
\frac{\Gamma\big(\tfrac{1-\nu_{-,l}}{2}\big)}{\Gamma\big(-\tfrac{\nu_{-,l}}{2}\big)} = 0.
\label{eq:quant}
\end{equation}
Any solution $E_l$ of this equation with $\Im E_l<0$ corresponds to a normalizable mode which decays to zero under time evolution. 

In the Hermitian case $V_0\to 0$, from Eq.~\eqref{eq:nus} it follows that $\nu_+=\nu_-\equiv\nu$, and Eq.~\eqref{eq:det0} requires either $D_\nu(0)=0$ or $D'_\nu(0)=0$. Taking into account Eq.~\eqref{eq:D0}, the former condition is satisfied whenever $\nu$ is an odd integer and the corresponding $\xi(x)$ is an odd function of $x$ (see Eq.~\eqref{eqS:xi}), while the latter is satisfied whenever $\nu$ is an even integer and $\xi(x)$ is an even function. Accordingly, we recover the standard spectrum of the harmonic oscillator with energy levels $E_n=\omega(n+1/2)$, with $n \in \mathbb{N}$.

Due to the imaginary potential, the Hamiltonian $H$ in Eq.~\eqref{eqS:H} is not Hermitean and therefore the left and right eigenstates generically differ. In particular, let us denote by $\{\xi_n\}$ the right eigenstates of $H$ with (possibly complex) eigenvalues $\{E_n\}$ and by $\{\tilde\xi_n\}$ the corresponding left eigenvectors, i.e., the right eigenvectors of $H^\dagger$. 
A generic time-dependent wavefunction $\psi(x,t)$ of the system can be expanded as
\begin{equation}
\psi(x,t)=\sum_{n} c_n\,\xi_n(x)\,e^{-iE_n t},\quad \mbox{where}\quad 
c_n=\langle\tilde\xi_n|\psi(t=0)\rangle.
\label{eq:modesum}
\end{equation}
Defining the overlap matrix
\begin{equation}
R_{mn}\equiv \int_0^\infty \!\!\dd x\, \xi_m(x)\,\xi^*_n(x),
\end{equation}
the first detection probability $F_n$ of the particle in $x>0$ reads
\begin{equation}
F_n =\int_0^\infty\!\! \dd x\,  |\psi(x,t=n\tau)|^2
=\sum_{m,k} c_m\,c_k^{*}\,R_{mk}\,e^{-i(\Omega_m-\Omega_k)n\tau}\,e^{-(\Gamma_m+\Gamma_k)n\tau},
\label{eq:Pexact}
\end{equation}
where  $\Omega_m$ and $\Gamma_m$ are the real and the opposite of the imaginary part of the energy $E_m=\Omega_m-i\Gamma_m$.

\begin{figure}
    \centering
    \includegraphics[width=0.5\linewidth]{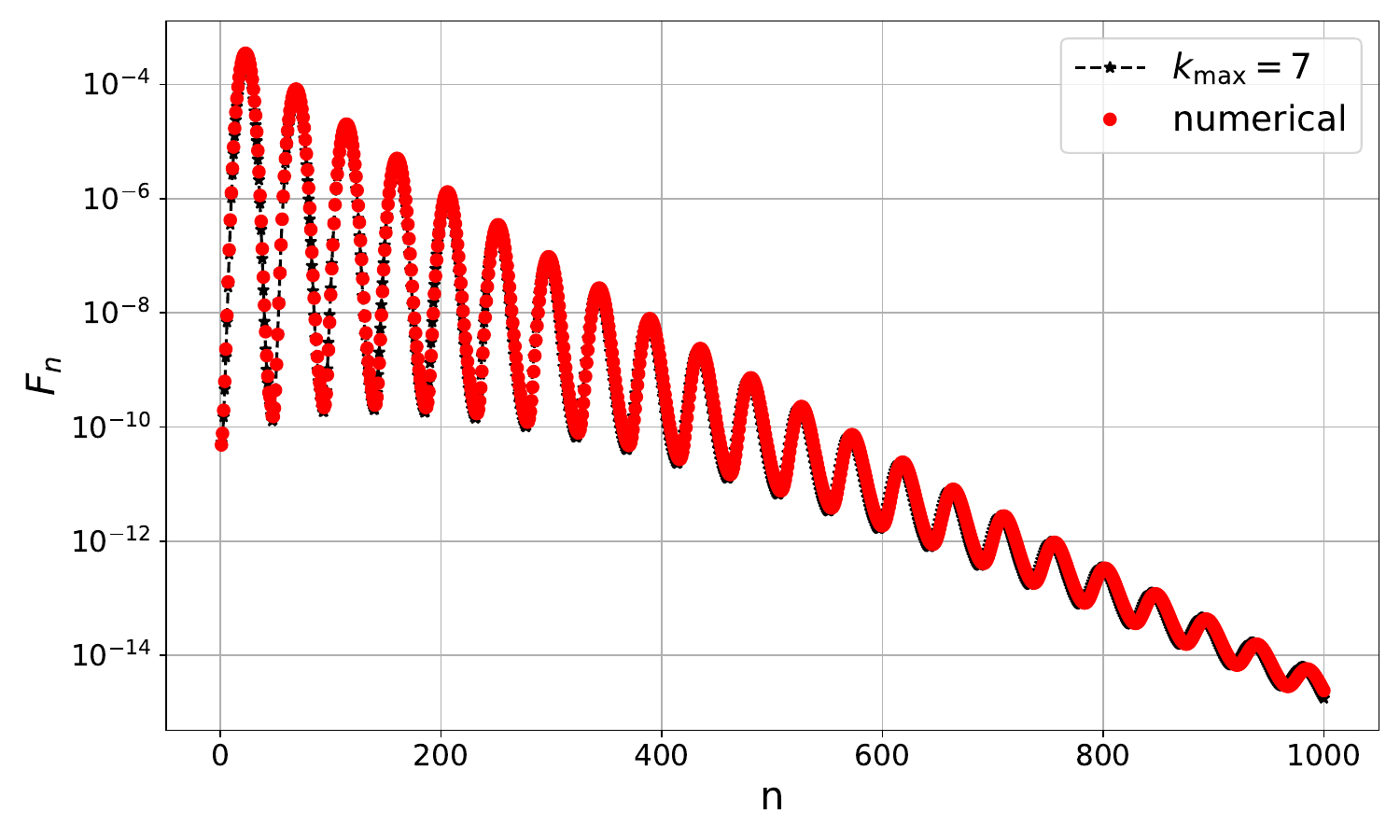}
    \caption{For an initial Gaussian wave function with $x_0 = - 0.9 \,$ and $\sigma = 0.3\,$ in a harmonic trap described by Eq.~\eqref{eqS:H}, with $\omega =7 \,$, the figure compares a fully numerical evaluation of $F_n$ with a numerical evaluation of Eq.~\eqref{eq:Pdiag}, where the summation is truncated at $k_{\text{max}} = 7$. The time interval $\tau = 0.01$} 
    \label{fig:num_vs_a}
\end{figure}

For strong absorption $V_0\gg 1$, the eigenmodes for $x\geq 0$ are strongly localized near $x=0$, and $R_{mn}$ becomes nearly diagonal.
Neglecting off-diagonal elements in Eq.~\eqref{eq:Pexact} yields
\begin{equation}
F_n \simeq \sum_k |c_k|^2\,R_{kk}\,e^{-2\Gamma_k n\tau}\quad \mbox{for}\quad V_0\gg 1.
\label{eq:Pdiag}
\end{equation}
At long times, the behavior of $F_n$ is thus determined by the $k_0$-th eigenstate which is characterized by the slowest temporal decay, i.e., 
\begin{equation}
F_n \simeq |c_{k_0}|^2\,R_{k_0k_0}\,e^{-2\Gamma_{k_0} n\tau},\quad \mbox{where}\quad 
\Gamma_{k_0}=\min_k \Gamma_k.
\end{equation}
In Fig.~\ref{fig:num_vs_a} we compare a fully numerical evaluation of $F_n$ (black solid line), obtained by evolving the initial wave packet described in the main text with the non-Hermitian Hamiltonian in Eq.~\eqref{eqS:H}, with a numerical evaluation of Eq.~\eqref{eq:Pdiag} (red dots). The two determinations display good agreement.

\section*{First Detection probability for a potential well}

We consider here the problem of determining the probability of first detection for a quanbtum particle in a potential well. Within the Zeno regime discussed in the main text, the problem can be conveniently studied in terms of the solutions of the one–dimensional stationary Schr\"odinger equation
\begin{equation}
\left[-\frac{1}{2}\frac{\dd^2}{\dd x^2}+\tilde{U}(x)\right]\psi(x;E)=E\,\psi(x;E),
\label{eqS:SchEq}
\end{equation}
where the potential $\tilde{U}$ includes the original potential well $U$ and the imaginary part of the potential which is introduced in the effective dynamics and which accounts for the back action of the measurements. 
Accordingly, the piecewise–constant potential $\tilde{U}$, consists of four regions,
\begin{equation}
\tilde{U}(x)=
\begin{cases}
0 & \quad\mbox{for}\quad x< -a/2 \quad \text{(region 1)},\\
- U_0 & \quad\mbox{for}\quad -a/2<x<0 \quad \text{(region 2)},\\
- U_0- iV_0, & \quad\mbox{for}\quad 0<x<a/2 \quad \text{(region 3)},\\
- iV_0, & \quad\mbox{for}\quad x>a/2 \quad \text{(region 4)},
\end{cases}
\qquad 
\label{eq:V-def}
\end{equation}
where $U_0>0$ and $V_0\ge 0$.
For a certain value of the spectral parameter $E$, we introduce the wave numbers
\begin{equation}
k_1=\sqrt{2E},\quad
k_2=\sqrt{2(E+V_0)},\quad
k_3=\sqrt{2(E+V_0+iV_m)}, \quad\mbox{and}\quad
k_4=\sqrt{2(E+iV_m)},
\label{eq:ks}
\end{equation}
corresponding, respectively, to regions 1, 2, 3, and 4. The choices of the branches of the square roots are done such that for scattering with  $\Re E>0$ one has $k_{1,2}\in\mathbb{R}$ and $\mathrm{Im}\,k_{3,4}\ge 0$, while for (quasi-)bound states, with $\Re E < 0$, the decaying branches are selected, such that $k_1=i\kappa_1$ with $\Re\,\kappa_1>0$, while $\mathrm{Im}\,k_4>0$ and $\mathrm{Im}\,k_3\ge 0$.

In each of the 4 regions introduced above, the corresponding expressions $\psi_1$, $\psi_2$, $\psi_3$, and $\psi_4$ of the stationary solution $\psi(x)$ of Eq.~\eqref{eqS:SchEq} are superpositions of plane waves, where
\begin{equation}
\psi_l(x) = A_l e^{ik_l x}+B_l e^{-ik_l x} \quad\mbox{with}\quad l = 1, 2, 3, 4,
\end{equation}
where the corresponding $k_l$ are given in Eq.~\eqref{eq:ks}. 
We focus on the case of left incidence in which the incoming scattering wave comes from the left. Accordingly, we impose
\begin{equation}
\psi_1(x)=e^{ik_1 x}+r\,e^{-ik_1 x}\quad\mbox{for}\quad x\to-\infty,\quad \mbox{while} \quad
\psi_4(x)=t\,e^{ik_4 x}\quad\mbox{for}\quad x\to+\infty,
\end{equation}
so that $A_1=1$, $B_1=r$, $A_4=t$, and $B_4=0$. 
Imposing the continuity of the solution $\psi(x)$ of Eq.~\eqref{eqS:SchEq} and of its first derivative $\psi'(x)$ at $x = - a/2$ by requiring $\psi_1(-a/2) = \psi_2(-a/2)$ and $\psi'_1(-a/2) = \psi'_2(-a/2)$ renders the conditions
\begin{align}
e^{-ik_1 a/2}+r\,e^{+ik_1 a/2}&=A_2 e^{-ik_2 a/2}+B_2 e^{+ik_2 a/2},\label{eq:L1}\\
ik_1\!\left(e^{-ik_1 a/2}-r\,e^{+ik_1 a/2}\right)&=ik_2\!\left(A_2 e^{-ik_2 a/2}-B_2 e^{+ik_2 a/2}\right).\label{eq:L2}
\end{align}
Solving Eqs.~\eqref{eq:L1}–\eqref{eq:L2} yields
\begin{equation}
\begin{pmatrix}A_2\\ B_2\end{pmatrix}
= S_{21}\begin{pmatrix}1\\ r\end{pmatrix},\quad\mbox{where}\quad
S_{21}\equiv \frac{1}{2}\begin{pmatrix}
\left(1+k_1/k_2\right)e^{i(k_2-k_1)a/2} & \left(1-k_1/k_2\right)e^{i(k_2+k_1)a/2}\\[7pt]
\left(1-k_1/k_2\right)e^{-i(k_2+k_1)a/2} & \left(1+k_1/k_2\right)e^{-i(k_2-k_1)a/2}
\end{pmatrix}.
\label{eq:S21}
\end{equation}
Similarly, the continuity of $\psi$ and $\psi'$ at $x=0$ requires $\psi_2(0) = \psi_3(0)$ and $\psi'_2(0) = \psi'_3(0)$, i.e.,
\begin{align}
A_2+B_2&=A_3+B_3,\label{eq:C1}\\
ik_2(A_2-B_2)&=ik_3(A_3-B_3).\label{eq:C2}
\end{align}
This can be written as
\begin{equation}
\begin{pmatrix}A_3\\ B_3\end{pmatrix}
= S_{32}\begin{pmatrix}A_2\\ B_2\end{pmatrix},\quad\mbox{where}\quad
S_{32}\equiv \frac{1}{2}\begin{pmatrix}
1+k_2/k_3 & 1-k_2/k_3\\[6pt]
1-k_2/k_3 & 1+k_2/k_3
\end{pmatrix}.
\label{eq:S32}
\end{equation}
Finally, imposing the continuity of the wavefunction and its derivative at $x = a/2$, i.e., 
$\psi_3(a/2) = \psi_4(a/2)$ and $\psi'_3(a/2) = \psi'_4(a/2)$, requires
\begin{align}
A_3 e^{+ik_3 a/2}+B_3 e^{-ik_3 a/2}&=t\,e^{+ik_4 a/2},\label{eq:R1}\\
ik_3\!\left(A_3 e^{+ik_3 a/2}-B_3 e^{-ik_3 a/2}\right)&=ik_4\,t\,e^{+ik_4 a/2}.\label{eq:R2}
\end{align}
This can be written as
\begin{equation}
\begin{pmatrix}1\\ k_4\end{pmatrix} t\,e^{+ik_4 a/2}
= S_{43} \begin{pmatrix}A_3\\ B_3\end{pmatrix},
\quad\mbox{where}\quad
S_{43}\equiv
\begin{pmatrix}
e^{+ik_3 a/2} & e^{-ik_3 a/2}\\
k_3 e^{+ik_3 a/2} & -\,k_3 e^{-ik_3 a/2}
\end{pmatrix}.
\label{eq:S43}
\end{equation}
Combining Eqs.~\eqref{eq:S21}, \eqref{eq:S32}, and \eqref{eq:S43},
we obtain
\begin{equation}
\begin{pmatrix}1\\ k_4\end{pmatrix} t\,e^{+ik_4 a/2}
= S_{43}\,S_{32}\,S_{21}\begin{pmatrix}1\\ r\end{pmatrix},
\label{eq:master}
\end{equation}
where $t$ and $r$ are the quantities to be determined. 
Solving this system of equations gives
\begin{equation}
t(E)=\frac{2ik_1\,e^{-ik_4 a/2}}{\ \Xi(E)\ }\quad\mbox{and}\quad
r(E)=-\frac{\Upsilon(E)}{\ \Xi(E)\ },
\label{eq:rt-final}
\end{equation}
where
\begin{equation}
\Xi(E)=\det\!\Big[\,M(E)\begin{pmatrix}1\\ ik_1\end{pmatrix},\ \begin{pmatrix}1\\ ik_4\end{pmatrix}\Big],
\qquad
\Upsilon(E)=\det\!\Big[\,M(E)\begin{pmatrix}1\\ -ik_1\end{pmatrix},\ \begin{pmatrix}1\\ ik_4\end{pmatrix}\Big],
\label{eq:XiU}
\end{equation}
and
\begin{equation}
M(E)=P(k_3)\,P(k_2), \quad\mbox{with}\quad
P(k)=
\begin{pmatrix}
\cos(ka/2) & \dfrac{\sin(ka/2)}{k}\\[6pt]
-\,k\sin(ka/2) & \cos(ka/2)
\end{pmatrix}.
\end{equation}
An explicit calculation of the determinants above yields the expressions
\begin{align}
\Xi(E)=&\ \cos\!\frac{k_3 a}{2}\,\cos\!\frac{k_2 a}{2}\,\Big(1+\frac{k_4}{k_1}\Big)
- i\,\cos\!\frac{k_3 a}{2}\,\sin\!\frac{k_2 a}{2}\,\Big(\frac{k_4}{k_2}+\frac{k_2}{k_1}\Big)\nonumber\\
&- i\,\sin\!\frac{k_3 a}{2}\,\cos\!\frac{k_2 a}{2}\,\Big(\frac{k_4}{k_3}+\frac{k_3}{k_1}\Big)
- \sin\!\frac{k_3 a}{2}\,\sin\!\frac{k_2 a}{2}\,\Big(\frac{k_3}{k_2}+\frac{k_2}{k_1}\frac{k_4}{k_3}\Big),
\label{eq:Xi-final}
\end{align}
and
\begin{align}
\Upsilon(E)=&\ \cos\!\frac{k_3 a}{2}\,\cos\!\frac{k_2 a}{2}\,\Big(1-\frac{k_4}{k_1}\Big)
- i\,\cos\!\frac{k_3 a}{2}\,\sin\!\frac{k_2 a}{2}\,\Big(\frac{k_4}{k_2}-\frac{k_2}{k_1}\Big)\nonumber\\
&- i\,\sin\!\frac{k_3 a}{2}\,\cos\!\frac{k_2 a}{2}\,\Big(\frac{k_4}{k_3}-\frac{k_3}{k_1}\Big)
- \sin\!\frac{k_3 a}{2}\,\sin\!\frac{k_2 a}{2}\,\Big(\frac{k_3}{k_2}-\frac{k_2}{k_1}\frac{k_4}{k_3}\Big).
\label{eq:Upsilon-final}
\end{align}
The presence of bound states is signaled by the condition $\Xi(E)=0$; equivalently, they correspond to poles of the scattering matrix.
In order to write the final state at time $t$, as reported in Eq.~(22) of the main text, let $\phi_E^{\mathrm{R},L}(x)$ and $\phi_E^{\mathrm{L},L}(x)$ denote the right- and left-scattering states for left incidence at real energies $E>0$, normalized such that
\begin{equation}
\braket{\phi_{E'}^{\mathrm{L},L}}{\phi_{E}^{\mathrm{R},L}}=\delta(E'-E)\,.
\end{equation}
For an initial packet $\psi_0(x)$ localized on the left,
\begin{equation}
\psi(x,t)=\sum_n c_n\,\phi_n^{\mathrm{R}}(x)\,e^{-iE_n t}
\;+\;\int_{0}^{\infty}\! dE\, c_L(E)\,\phi_E^{\mathrm{R},L}(x)\,e^{-iEt},
\label{eq:time-evol}
\end{equation}
with biorthogonal coefficients
\begin{equation}
c_n=\braket{\phi_n^{\mathrm{L}}}{\psi_0}\,,\qquad
c_L(E)=\braket{\phi_E^{\mathrm{L},L}}{\psi_0}\,.
\end{equation}
This provides a proof of Eq.~(22) of the main text; moreover, by applying the definition in Eq.~(4), one can evaluate $F_n$ for this example.

\section*{First-passage-time statistics of the Wiener process subject to stroboscopic measurements}

For comparison with the quantum results of the main text, it is useful to recall the corresponding behavior in the classical Brownian motion, for which we take the Wiener process as a representative example. This process is characterized by the propagator
\begin{equation}
    G(x,t|x_{0},0)
    =\frac{1}{\sqrt{4\pi D t}}
    \exp\!\left[-\frac{(x-x_{0})^{2}}{4Dt}\right].
\end{equation}
The probability density $p_{x_0}(t)$ for the first passage time to the origin of the Wiener process which starts at position $x_0$ is given by~\cite{Redner2001}
\begin{equation}
p_{x_0}(t) =  \frac{|x_0|}{\sqrt{4\pi D t^3}}\exp[- x_0^2/(4 D t)].
\label{eqS:fistdet}
\end{equation}
Accordingly, at long times $t \gg x_0^2/(2 D t)$, $p_{x_0}(t)$ decays $\propto t^{-3/2}$, independently of $x_0$. 

In the quantum problem discussed in the main text, the dynamics is probed stroboscopically. In fact, a continuous monitoring --- such as that implied in the classical problem of first passage --- would freeze the evolution through the quantum Zeno effect. A natural question is therefore whether the discretization of the observation times affects the universal scaling exponent in the classical case. As we show below, it does not. For sufficiently large $t/\tau$, one intuitively expects that restricting the observation of the process to multiples of $\tau$ should not modify the long-time statistics.

\begin{figure}
    \centering
    \includegraphics[width=0.5\linewidth]{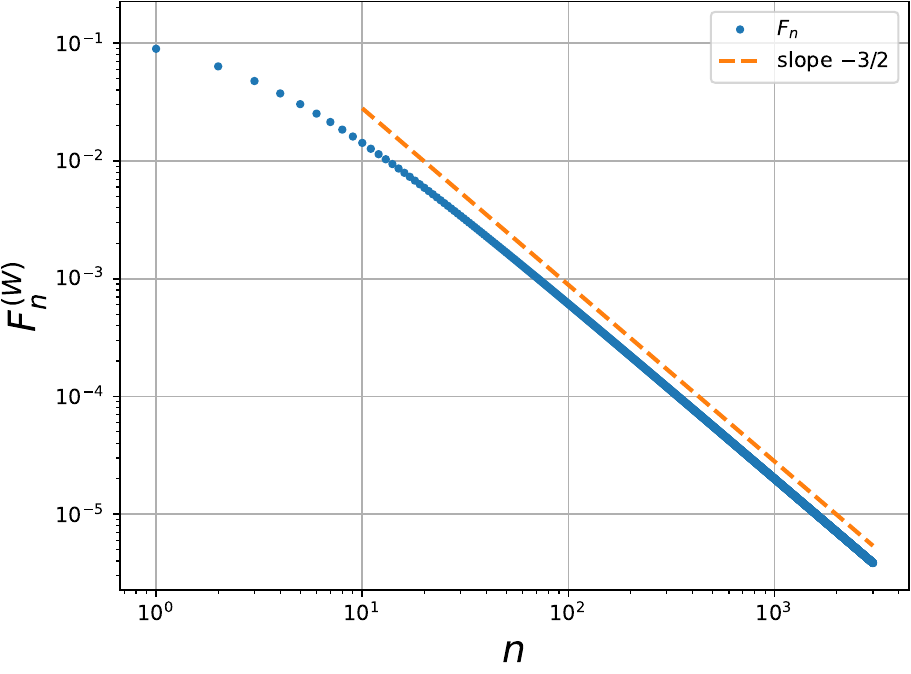}
    \caption{First detection probability $F_n^{(W)}$ at time $n\tau$ for the classical Wiener process observed at stroboscopic times with period $\tau$. The dashed line corresponds to the decay $F_n^{(W)} \simeq n^{-3/2}$.
    } 
    \label{fig:plot_classical}
\end{figure}

In particular, for the Wiener process the probability density $P(\{x_i,t_i\}_{i=1,\dots,n})$ to observe the process at position $x_i$ at time $t_i$, with $i=1,\ldots, n$ is given by
\begin{equation}
P(\{x_i,t_i\}_{i=1,\dots,n}) = G(x_n,t_n|x_{n-1},t_{n-1}) G(x_{n-1},t_{n-1}|x_{n-2},t_{n-2})\cdots G(x_1,t_1|x_0,0),
\end{equation}
where we assumed that the process starts from position $x_0$ at time $t=0$. Accordingly, the probability $F_n^{(W)}$ to find the Wiener process at $x>0$ for the first time at $t_n = n\tau$ after that in all the previous observations at times $t_i = i\tau$, with $i =1, \ldots, n-1$ it has been found at $x<0$ is given by
\begin{equation}
    F_n^{(W)}=\int_0^{\infty}\!dx_n
    \int_{-\infty}^0\!dx_{n-1}\cdots
    \int_{-\infty}^0\!dx_1\;
    G(x_n,n\tau|x_{n-1},(n-1)\tau) G(x_{n-1},(n-1)\tau|x_{n-2},(n-2)\tau)\cdots G(x_1,\tau|x_0,0). 
\end{equation}
After a convenient change of variable $x_i = y_i \sqrt{2 D\tau}$, $F_n$ can be written as 
\begin{equation}
    F_n^{(W)}=\frac{1}{(2\pi)^{n/2}}\int_0^{\infty}\!dy_n
    \int_{-\infty}^0\!dy_{n-1}\cdots
    \int_{-\infty}^0\!dy_1\;
    \exp\left\{-\sum_{i=1}^n (y_i-y_{i-1})^2/2\right\} \quad\mbox{where}\quad y_0 \equiv  x_0/\sqrt{2 D\tau}.
\end{equation}
A numerical evaluation of $F_n^{(W)}$ confirms that, for sufficiently large $n$, it decays algebraically as a function of $n$ with 
\begin{equation}
    F_n^{(W)} \simeq n^{-3/2},
\end{equation}
consistently with the behavior in Eq.~\eqref{eqS:fistdet} for the continuous observation, using $t=n\tau$.
This is shown in Fig.~\ref{fig:plot_classical}, where  $F_n^{(W)}$ is plotted as a function of $n$, for $y_0=-1$,
and where the dashed line correspond to an algebraic decay with exponent $-3/2$.

%

\bibliographystyle{apsrev4-2}